\documentclass[twocolumn,letterpaper,aps,prl,superscriptaddress,showpacs,floatfix]{revtex4-1}

\usepackage{amsmath}
\usepackage{graphicx}	

\usepackage{xspace}	

\newcommand{\sqsn}{\mbox{$\sqrt{s_{_{NN}}}$}\xspace}
\newcommand{\sqs}{\mbox{$\sqrt{s}$}\xspace}

\begin{document}

%
\title{ Has the QCD critical point been observed at RHIC? \\
- A Rebuttal }

%
%
%
\author{ Roy A. Lacey}
\email[E-mail: ]{Roy.Lacey@Stonybrook.edu}
\affiliation{Department of Chemistry, 
Stony Brook University, 
Stony Brook, NY, 11794-3400, USA}
\affiliation{Dept. of Physics, 
Stony Brook University, 
Stony Brook, NY, 11794, USA}

\date{\today}

\begin{abstract}
This note rebuts an old but recurring claim by Antoniou, Davis and Diakonos \cite{Antoniou:2016hbx}, that the critical point and associated critical 
exponents reported in Ref. \cite{Lacey:2014wqa}, is based on an erroneous treatment of scaling relations near the critical point.

\end{abstract}

	



\maketitle
 

%
As indicated in Ref. \cite{Lacey:2014wqa}, we exploit the phenomenology of thermal models 
to establish the chemical freeze-out region (assumed to be close to the coexistence region), and associate the combinations 
of freeze out temperature and baryon chemical potential ($T, \mu_B$) with the values 
for \sqsn [see for example, Refs. \cite{Cleymans:2006qe,Becattini:2012sq,Andronic:2009qf}]. This means that;
\begin{itemize}
\item  	\sqsn  is used as a control parameter to measure the "distance" $\tau_{\sqs}$ to the CEP, where 
$\tau_{\sqs} = (\sqsn - \sqsn^{\text{cep}})/\sqsn^{\text{cep}}$. Here it is important to emphasize that Finite Size Scaling is a very general 
and flexible technique that does not require precise nor detailed information about the direction of approach to the CEP in the ($T, \mu_B$)-plane. 
A variable that can give a reasonable measure of the distance to the CEP will suffice.

\item 	Our Finite Size Scaling analysis is actually performed for fixed values of  $T$ (and $\mu_B$) at each collision energy. 
This is validated by recent extractions of  $T$ and  $\mu_B$ (at chemical freeze out) as a function of collision centrality or 
system size (see Fig.1) \cite{Das:2014oca}. Thus, the relevant critical exponent in our Finite Size Scaling analysis 
is $\gamma$,  {\bf not} $\alpha$, as claimed by the authors of Ref. \cite{Antoniou:2016hbx}. 
Note as well, that for isothermal freeze out, the isentropic compressibility $\kappa_S$ can be 
expressed in terms of the isothermal compressibility $\kappa_T$ and the ratio $C_P/C_V$ of the 
constant pressure ($P$) and constant volume ($V$) heat capacities ($C$).

%
%
\begin{figure}[ht]
\includegraphics[width=1.0\linewidth]{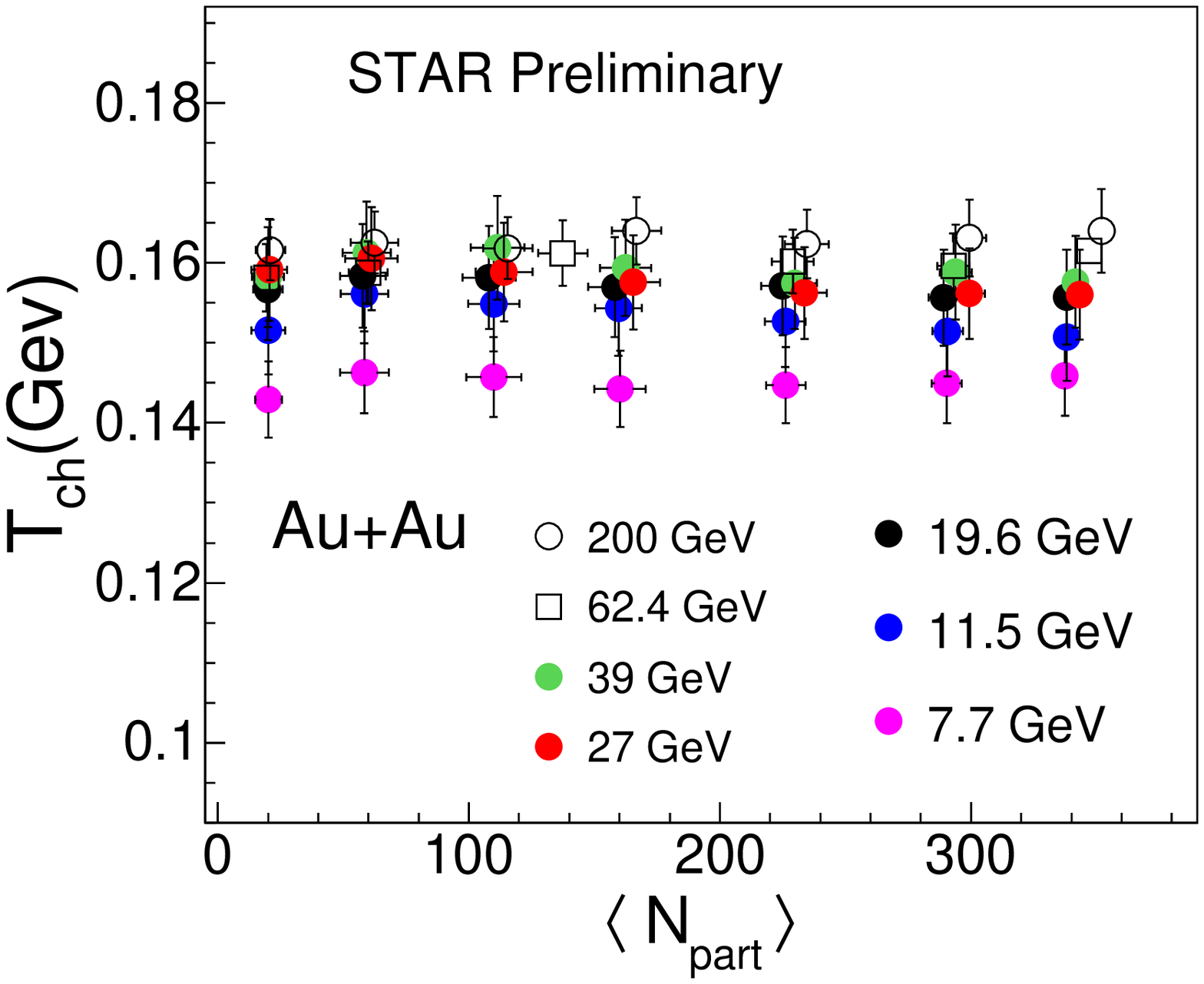}
\caption{Chemical freeze out temperature vs. $\left< N_{\text{part}} \right>$ or collision centrality \cite{Das:2014oca}.
}
\label{Fig1}
\end{figure}
\end{itemize}

Our Finite-Size Scaling analysis is further validated in Fig. 5 of  Ref. \cite{Lacey:2014wqa}, where it is shown that 
the extracted critical exponents, coupled with the estimated location of the CEP ($T^{cep}$ and $\mu_B^{cep}$) , {\bf do} 
result in the requisite collapse of the full data 
set on to a single curve or {\em scaling function}. This constitutes a crucial and compelling ``closure'' test of the efficacy of 
our Finite Size Scaling analysis and the associated parameters extracted. Incidentally, the estimates given for the CEP and the 
critical exponents in Ref. \cite{Lacey:2014wqa} have been recently validated \cite{Lacey:2016tsw} with a different 
set of observables.

\section*{Acknowledgments}
This research is supported by the US DOE under contract DE-FG02-87ER40331.A008.

\bibliography{ref_rebut}   

\end{document}